\documentclass[twocolumn,showpacs,preprintnumbers,aps]{revtex4}
\usepackage{graphicx} % Include figure files
\usepackage{dcolumn}  % Align table columns on decimal point
\usepackage{bm}       % bold math

\begin{document}

\title{Quasiperiodic Bloch-like states in a surface wave experiment}

\author{M.  Torres\footnote{Electronic address: manolo@iec.csic.es}}
\affiliation{Instituto de F\'{\i}sica Aplicada, Consejo Superior de
  Investigaciones Cient\'{\i}ficas, Serrano 144, 28006 Madrid, Spain.}

\author{J.L. Arag\'on\footnote{Electronic address:
    aragon@fata.unam.mx}} \affiliation{Centro de F\'{\i}sica Aplicada
  y Tecnolog\'{\i}a Avanzada, Universidad Nacional Aut\'onoma de
  M\'exico, Apartado Postal 1-1010, Quer\'etaro 76000, M\'exico.}

\author{J. P. Adrados} \affiliation{Instituto de F\'{\i}sica
  Aplicada, Consejo Superior de Investigaciones Cient\'{\i}ficas,
  Serrano 144, 28006 Madrid, Spain.}

\author{P. Cobo} \affiliation{Instituto de Ac\'ustica, Consejo
  Superior de Investigaciones Cient\'{\i}ficas, Serrano 144, 28006
  Madrid, Spain.}

\author{S. Tehuacanero} \affiliation{Instituto de F\'{\i}sica,
  Universidad Nacional Aut\'onoma de M\'exico, Apartado Postal 20-364,
  M\'exico 01000, Distrito Federal.}

\date{\today}

\begin{abstract}
  Bloch-like surface waves associated with a quasiperiodic structure
  are observed in a classic wave propagation experiment which consists
  of pulse propagation with a shallow fluid covering a
  quasiperiodically drilled bottom. We show that a transversal pulse
  propagates as a plane wave with quasiperiodic modulation, displaying
  the characteristic undulatory propagation in this quasiperiodic
  system and reinforcing the idea that analogous concepts to Bloch
  functions can be applied to quasicrystals under certain
  circumstances.
\end{abstract}


\maketitle

The main difficulty towards the development of a systematic analytic
approach to the transport properties of quasiperiodic systems has been
the absence of an analogous Bloch theorem approach as used in the
periodic case. In the first efforts to apply a modified version of the
Bloch theorem, it was noticed that the dense spectrum of quasiperiodic
systems is dominated by only a few special reciprocal lattice points
that may be taken to construct a quasi-Brillouin zone \cite{Smith}.
Thus, by considering only the dominant Fourier components, the atomic
distribution can be expanded in terms of a discrete aperiodic lattice.
Wave functions of the form $\Psi _\mathbf{k} = u_\mathbf{k}
(\mathbf{r}) e^{i \mathbf{k} \cdot \mathbf{r}}$ will therefore solve
the Schrodinger equation. In this case $u_\mathbf{k}$ is quasiperiodic
and should formally be defined on a countable dense set of reciprocal
lattice vectors. But, by the above considerations, this expansion is
useful since the Fourier development of the modulation function
$u_\mathbf{k}$ can be restricted to the few special reciprocal vectors
that dominate the spectra. Thus, Bloch-like states could describe the
plane wave propagation in so schematized quasicrystals and
free-electron-like bands are expected. Recently this idea was
experimentally tested showing that analogous concepts to Bloch
functions can be applied to quasicrystals \cite{Rotenberg}.

The classic wave propagation in quasicrystalline systems was addressed
in a first seminal acoustic experiment of He and Maynard \cite{He} by
the feature that acoustical waves are ideal tools to investigate
formally similar quantum propagation effects \cite{Maynard}. On the
other hand, appearance of the quasicrystalline symmetry in fluids
dynamics was firstly predicted theoretically by Zaslavsky and
co-workers \cite{Zaslavsky1} and a simulation similar to the
conditions of the present experiments was reported in Ref.
\cite{Beloshapkin}.  Finally, compressible quasisymmetric flows were
considered in Ref. \cite{Morgulis}, whereas a general outlook on order
and disorder in fluid motion can be found in the experiments of Gollub
\cite{Gollub}.

In this letter we shall see that a discrete restricted spectral
scenario can be displayed by means of impulsive waves in hydrodynamic
quasicrystals, where we observe Bloch-like surface waves.  The waves
are generated at the frequencies corresponding to the Fourier
components of the quasiperiodic structure at the dominant diffraction
spots. The observed Bloch-like waves are plane waves with
quasiperiodic modulation generated when a pulse propagates
transversally to the quasiperiodic structure. Liquid surface waves
shape a quasiperiodic grid that obeys the so-called Octonacci
sequence, previously studied \cite{Sire,Borcherds} but never observed
in any experiment.

The quasiperiodic structure involved in our experiment is the
octagonal Ammann-Beenker tiling composed by squares and rhombuses
\cite{Senechal}. Associated with this octagonal tiling is a
quasiperiodic sequence, named the Octonacci sequence \cite{Sire}, that
can be generated starting from two steps $L$ and $S$, which are
related according to the irrational ratio $L / S = 1 + \sqrt{2}$, by
iteration of substitution rules: $L \rightarrow LSL$ and $S
\rightarrow L$.

The experiments are performed with surface waves generated on a
shallow fluid that covers the quasiperiodically drilled bottom of a
transparent vessel. Such experiments are similar to others realized in
vessels with periodic bathymetry and described elsewhere
\cite{Torres1,Torres2} but here not only a continuous wave excitation
driven by a vertical monofrequency vibration but also a new experiment
of transversal pulse propagation is performed. The bottom dimples are
located at $121$ vertices of the octagonal tiling. The edge length $l$
of the tiling is $8$ mm with an error lower than $0.4$\%, the radius
$r$ of the cylindrical bottom wells is $1.75$ mm with an error lower
than $1$\% and their depth $d$ is $2$ mm.  The depth of the liquid
layer over the cylindrical wells is given by $h_2 = h_1 + d$, where
$h_1$ is the depth of the thin liquid layer covering the bottom of the
vessel among holes.

\begin{figure}[!hb]
  \includegraphics[width=7.0cm]{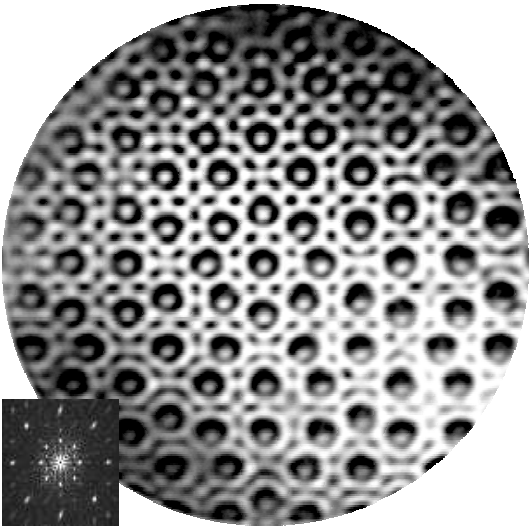}
\caption{Snapshot of the system vibrating vertically at a
  frequency of $35$ Hz, under conditions of continuous wave excitation.
  The inset shows its Fourier spectrum with a well defined and
  discrete set of relevant components.}
 \label{fig:fig1}
\end{figure}

Under conditions of continuous-wave excitation, an inertial
hydrodynamical undulatory instability grows over the bottom wells when
the system vibrates vertically at a frequency of $35$ Hz.  Such an
instability becomes remarkable (Fig. \ref{fig:fig1}) due to the high
density and the very low surface tension of the liquid \cite{Liquid}.
Oscillating bulges over dimples are connected by surface waves with
shorter wavelength that decorate the shallow liquid region among holes
with $h_1$ being $0.4$ mm. This is a physical scenario similar to that
of the Kronig-Penney model but adapted here to a $2$D quasiperiodic
system. It should be remarked that Fig. \ref{fig:fig1} is the first
available experimental example of a quasiperiodic pattern of waves not
arising from a non-linear instability such as the Faraday instability.
Standing waves are coupled to only two rings of Fourier wave
components of the quasiperiodic bottom structure, as shown in the
inset of Fig.  \ref{fig:fig1}. Their wavenumber ratio is $1 +
\sqrt{2}$, as it can be measured in the diffraction pattern (see
below). If the effective Fourier transform of a quasiperiodic
structure is restricted to a discrete set of Fourier peaks as in this
case, then Bloch-like modulation functions can be used to describe the
wave propagation in such a simplified quasicrystal \cite{Janot}.  The
pattern of Fig. \ref{fig:fig1} is due to the strong coupling between
liquid surface waves and the bottom quasiperiodic topography and does
not depend on the shape of the vessel boundary. Undistinguishable
patterns are generated with octagonal or circular boundaries. Although
the boundaries are reflecting vertical walls, the boundary symmetry
matches the symmetry of this standing wave experiment allowing the
mentioned boundary-independent strong wave coupling. Due to the
accuracy on realizing the setup, localization phenomena do not appear
in the described experiment. However, slight tiltings of the vessel
generate wave domains \cite{Torres1,Torres2}; furthermore, point and
linear defects can be easily introduced in the system by dropping
mercury on the bottom dimples to study new interesting wave
localization phenomena \cite{Torres3}.

\begin{figure}[!hb]
  \includegraphics[width=8.0cm]{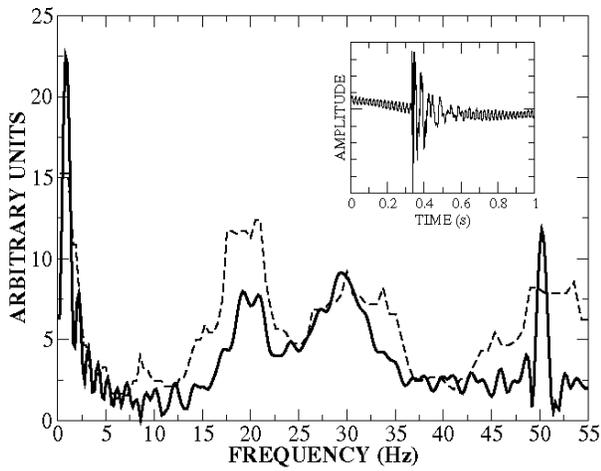}
\caption{The solid line is the Fourier transform of the impulsive
  signal as picked up by an accelerometer at the center of the vessel.
  It shows clear resonances at approximately $20$, $30$ and $50$ Hz.
  Such resonances are specific of the system.  The first peak at very
  low frequency, close to the origin, corresponds to the Fourier
  transform of the square pulse that excites the system. The signal in
  the time domain is presented in the inset. The dashed line is the
  gray scale of the subpattern along the $\Gamma-X$ direction as
  represented in Fig.  \ref{fig:fig4} versus frequency. The gray scale
  (between $0$ and $1$) was rescaled to match the peak of the Fourier
  transform at $30$ Hz. Three standing waves indicating narrow band
  gaps appear at approximately $20$, $32$ and $50$ Hz.}
 \label{fig:fig2}
\end{figure}

The propagation of a plane wave through the octagonal quasiperiodic
structure can be visualized by means of a experiment of wave pulse
propagation. A coupling between the vertical waves of the
vessel-liquid system and the surface waves generated by a transversal
pulse is expected, modelling in this way the propagation of a plane
wave through the quasiperiodic structure.  This experiment was
realized in a vessel with octagonal boundary. The octagon side $L$ is
$4$ cm and it is perpendicular to the $\Gamma-X$ direction of the well
structure. The surface ratio $f$ between bottom holes and the whole
octagon is about $0.15$ with $h_1$ being $0.5$ mm. The system is
excited near the octagonal boundary with a wave pulse parallelly to
the liquid surface and perpendicularly to a side of the octagon. The
signal is picked up by means of a Br\"uel \& Kjaer 4344 accelerometer
placed at the center of the vessel and it is processed by means of a
digital acquisition system. The impulsive signal and the corresponding
Fourier transform are shown in Fig. \ref{fig:fig2} (inset and solid
line, respectively). Three clear spectral peaks of the vibrational
vessel-liquid system appear at about $20$, $30$ and $50$ Hz. As we
shall see, such resonances indicate the existence of three narrow band
gaps in the liquid surface wave propagation \cite{Torres2},
\textit{i.e}. standing liquid waves are generated at approximately the
above generated frequencies.

\begin{figure}[!ht]
  \includegraphics[width=7.0cm]{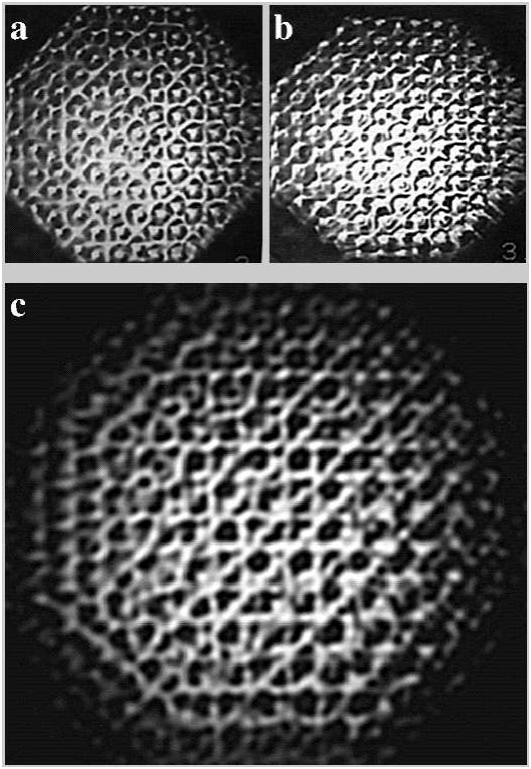}
\caption{Temporal sequence of patterns observed when the system is
  disturbed with a transverse pulse. (a) Quasicrystalline pattern
  observed at the start of each pulse. (b) A transitory weak
  turbulence is observed after $0.04$ s. (c) Standing waves draw clear
  quasiperiodic grids between $0.08$ and $0.24$ s. This figure should be
  looked diagonally at grazing incidence.}
 \label{fig:fig3}
\end{figure}

At the start of each pulse the liquid \textit{feels} the perturbation
and a nice quasicrystalline surface wave pattern suddenly appears
[Fig. \ref{fig:fig3}(a)]. A transitory weak turbulence arises in the
system after scarcely $0.04$ s [Fig.  \ref{fig:fig3}(b)], whereas
robust standing waves drawing clear quasiperiodic grids can be
observed between $0.08$ and $0.24$ s on the liquid surface [Fig.
\ref{fig:fig3}(c)]. Finally, quasiperiodic grid patterns decay until
the arrival of the next pulse. As wave phase velocities are about $11$
cm s$^{-1}$ and the wave group velocity is nearly null near the gaps,
times for an echo at the boundaries to come back are much longer than
observation time.

\begin{figure}[!ht]
  \includegraphics[width=7.5cm]{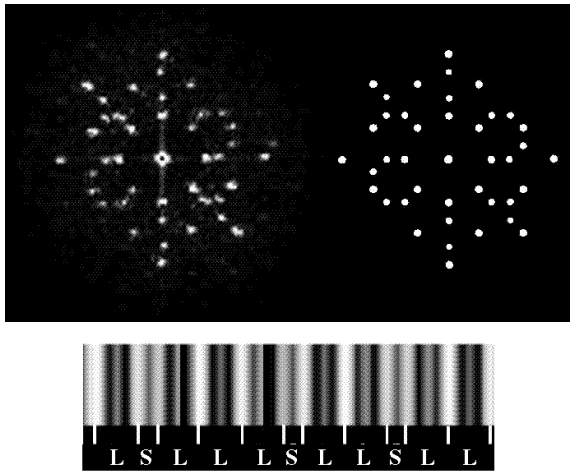}
\caption{Experimental (top left) and theoretical (top right) Fourier
  transform of the pattern shown in Fig. \ref{fig:fig3}(c). At the
  bottom, the inverse Fourier transform of a subpattern along the
  $\Gamma-X$ is shown. The Octonacci sequence is clearly recovered.}
 \label{fig:fig4}
\end{figure}

The robust quasiperiodically spaced standing waves shown in [Fig.
\ref{fig:fig3}(c)] are generated by discrete Bragg resonances and thus
can be considered quasiperiodic Bloch-like waves. To verify this,
first note that the irrational ratio $LS / L = \sqrt{2}$ is apparent
in our experiment [Fig. \ref{fig:fig3}(c)]. Now, using a
crystallography-oriented computer program \cite{Hovmuller}, the
Fourier transform of the pattern of Fig. \ref{fig:fig3}(c) is
calculated and shown in Fig.  \ref{fig:fig4} (top left). Such
diffraction pattern matches with an adequate subset of the diffraction
pattern of the direct product of both orthogonal Octonacci sequences
calculated according to theoretical methods \cite{Sire,Aragon} as
shown in Fig.  \ref{fig:fig4} (top right). The absence of some
diffraction peaks indicates the directional character of the impulsive
action. The pulse runs along the $\Gamma-X$ direction from the upper
left to the bottom right corner in both patterns at the top of Fig.
\ref{fig:fig4}.  Along this direction, a intensity profile is taken in
the experimental pattern and recovered the inverse Fourier transform
of that unidimensional diffraction subset. The result is shown in Fig.
\ref{fig:fig4} (bottom) which displays an Octonacci sequence, and it
matches with that generated theoretically starting from the above
mentioned substitution rules. Finally, the above described intensity
profile along the $\Gamma-X$ direction is scaled according to the
wavenumber of the waves of the diffraction pattern shown in Fig.
\ref{fig:fig3}(c). Such scale is then changed according to the
approximate dispersion relationship given \cite{Torres2} by
\[
\omega ^2 = gk \left( 1 + \frac{T}{\rho g} k^2 \right) \tanh (k
h_0),
\]
where $h_0 = h_1 (1-f) + h_2 f$, $\omega$ is the angular frequency,
$k$ is the wave number, $g$ is the acceleration due to gravity, $T$ is
the liquid surface tension and $\rho$ is the liquid density
\cite{Liquid}. In Fig. \ref{fig:fig2} (dashed lines) the gray scale
intensity profile is plotted versus the frequency according to the
above mentioned change of scale. The first maximum is scaled by the
wavenumber $k = 11$ cm$^{-1}$, that corresponds to the main ubiquitous
wave appearing in the experimental pattern of Fig.  \ref{fig:fig3}(c).
As it can be seen, diffraction peaks which represent narrow band gaps
closely match in frequency with those independently measured also in
Fig.  \ref{fig:fig2}. Thus, in this restricted scenario, where the
resonances of the vibrational coupling generates a discrete spectrum,
the wave pattern observed in Fig.  \ref{fig:fig3}(c) corresponds to
quasiperiodic Bloch-like states.

If experiments of pulse propagation are realized in vessels with
periodic bathymetry \cite{Torres1,Torres2} no signal of turbulence
appears. Thus, as remarked in a different context
\cite{Borcherds,Zaslavsky2}, the quasiperiodicity of the
hydrodynamical system could be the origin of the weak chaos
observed in the described experiments just at the start of pulses,
when amplitudes are higher and hence the nonlinearity is stronger.
Then, a rapidly increasing number of incommensurable Fourier
harmonics can grow due to the finite frequency bandwidth of the
pulse and the incommensurate nature of the system. This gives rise
to the pre-turbulent state of the surface waves. When the
multiscattering becomes weaker, the Fourier mode cascade decays
and the propagative wave exhibits a clean quasiperiodic grid
pattern.

\begin{figure}[!ht]
  \includegraphics[width=7.5cm]{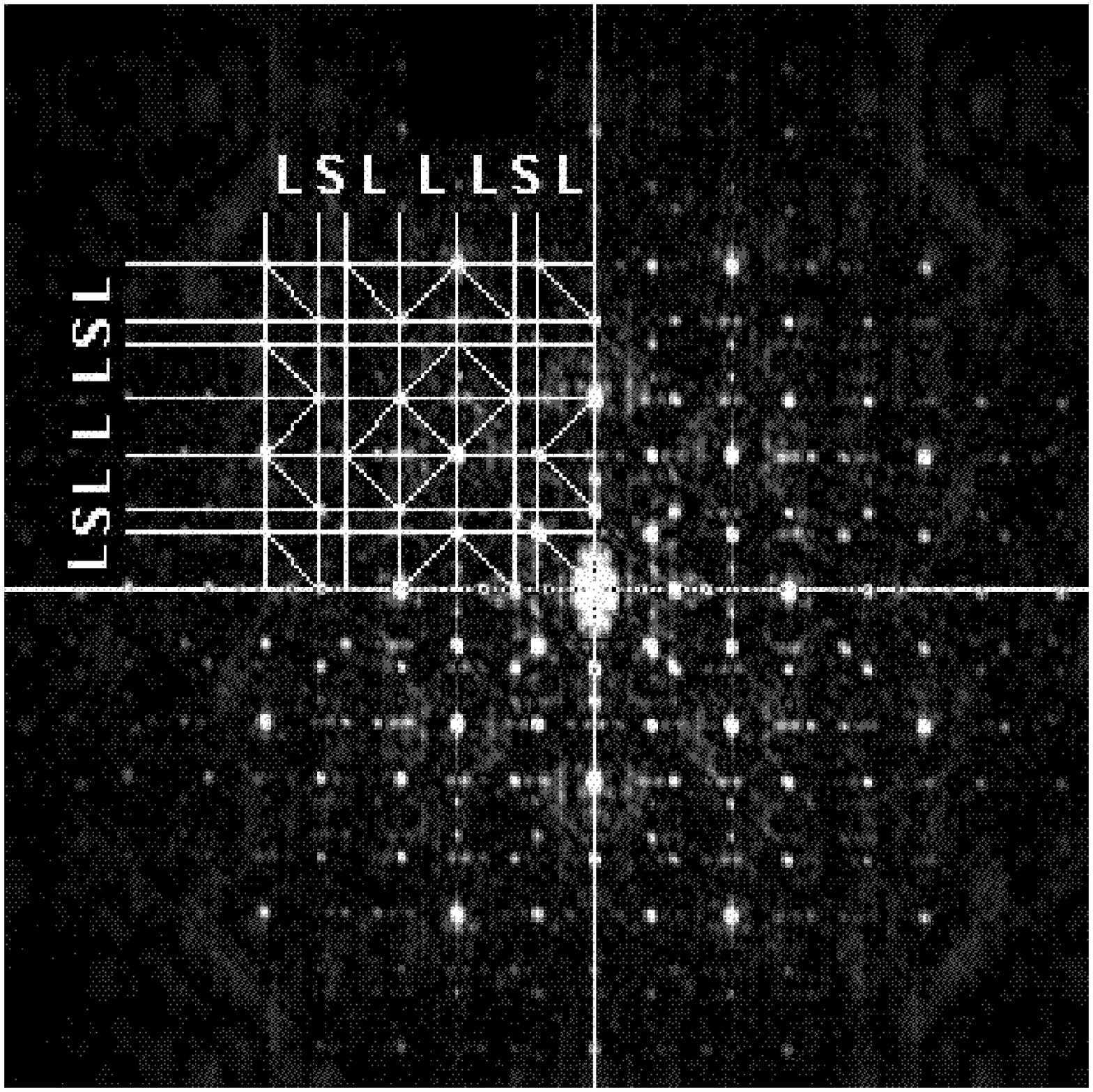}
\caption{Inverse Fourier transform of pattern shown in
  Fig. \ref{fig:fig3}(b). The structure of the octagonal tiling
  underlying on the well quasiperiodic arrangement of the vessel
  bottom is recovered.}
 \label{fig:fig5}
\end{figure}

Anyway, the quasiperiodic structure underlies in spite of the weak
turbulence apparent in Fig. \ref{fig:fig3}(b), which must be
looked at grazing incidence to recognize Octonacci quasiperiodic
sequences. This is evident in Fig. \ref{fig:fig5}, which is the
inverse Fourier transform of Fig. \ref{fig:fig3}(b). The direct
product of two orthogonal Octonacci sequences is recovered there,
showing a patch of the well known octagonal tiling filled with
square and rhombic tiles \cite{Senechal}.

In conclusion, we have shown Bloch-like surface waves associated with
a quasiperiodic structure in a classic wave propagation experiment.
These waves draw clear quasiperiodic grids that obey the Octonacci
sequence. Our results along with earlier ones \cite{He} can be helpful
to understand the characteristic undulatory propagation in
quasiperiodic systems.

\begin{acknowledgments}
 This work has been supported by MCYT  (Project No.BFM20010202)
  and DGAPA-UNAM (Proyect 108199).
\end{acknowledgments}

\newpage

\begin{thebibliography}{10}
\bibitem{Liquid} Fluorinert FC75 of the 3M Company, a fully
  fluorinated organic liquid with a low kinematic viscosity = 0.8
  10$^{\rm{}2}$ cm$^{\rm{}2}$/s, lower than that of the water, very
  low surface tension T = 15 dyn/cm and very high density = 1.77
  g/cm$^{\rm{}3}$.
\end{thebibliography}
\end{document}